# Adaptive Regularized Low-Rank Tensor Decomposition for Hyperspectral Image Denoising and Destriping

Dongyi Li, Dong Chu, Xiaobin Guan, Wei He, *Member, IEEE*, Huanfeng Shen, *Senior Member, IEEE*

*This work has been submitted to the IEEE for possible publication. Copyright may be transferred without notice, after which this version may no longer be accessible.*

*Abstract*—Hyperspectral images (HSIs) are inevitably degraded by a mixture of various types of noise, such as Gaussian noise, impulse noise, stripe noise, and dead pixels, which greatly limits the subsequent applications. Although various denoising methods have already been developed, accurately recovering the spatial-spectral structure of HSIs remains a challenging problem to be addressed. Furthermore, serious stripe noise, which is common in real HSIs, is still not fully separated by the previous models. In this paper, we propose an adaptive hyper-Laplacian regularized low-rank tensor decomposition (LRTDAHL) method for HSI denoising and destriping. On the one hand, the stripe noise is separately modeled by the tensor decomposition, which can effectively encode the spatial-spectral correlation of the stripe noise. On the other hand, adaptive hyper-Laplacian spatial-spectral regularization is introduced to represent the distribution structure of different HSI gradient data by adaptively estimating the optimal hyper-Laplacian parameter, which can reduce the spatial information loss and over-smoothing caused by the previous total variation regularization. The proposed model is solved using the alternating direction method of multipliers (ADMM) algorithm. Extensive simulation and real-data experiments all demonstrate the effectiveness and superiority of the proposed method.

*Index Terms*—Hyperspectral images (HSIs), adaptive hyper-Laplacian, tensor completion, denoising and destriping.

## I. INTRODUCTION

Hyperspectral images (HSIs) obtained by imaging spectroscopy contain hundreds of spectral bands, providing rich spectral information. Taking advantage of the abundant spatial and spectral information, HSIs have been widely applied in various applications, such as mineral exploration, precision farming, environmental protection [1], [2], and so on. However, the narrower spectral channels usually mean that the sensors receive fewer signals, making HSIs highly susceptible to interference from various types of noise, such as stripe noise, impulse noise, Gaussian noise, dead lines, and dead pixels [3], [4]. Consequently, developing effective HSI denoising methods is a challenging and necessary task for the various subsequent applications.

In recent decades, considerable progress has been made in the field of HSI denoising. The straightforward approach is to treat each band of the HSI as a two-dimensional (2-D) grayscale image and then adopt a traditional grayscale image denoising method [5], such as block-matching and three-dimensional (3-D) filtering (BM3D) [6] or weighted nuclear norm minimization (WNNM) [7]. However, these approaches obtain relatively poor restoration performances since they ignore the strong spectral correlation of HSIs. Machine learning and variational-based methods are the commonly used methods that can use the spectral correlation of HSIs. Due to the formidable nonlinear fitting capability of deep learning, a number of studies have integrated deep learning into HSI denoising [8-15], However, due to the complex interplay of the multiple factors contributing to the degradation of real-world HSIs, the current deep learning based methods for HSI denoising can struggle to accurately replicate this intricate physical process. Variational-based methods can efficiently integrate the various prior knowledge of the HSI and noise, allowing for a more flexible adaptation to different noise scenarios. Recently, HSI denoising methods based on the variational framework considering the spectral correlation of HSIs have emerged as mainstream techniques, mainly including low-rank matrix decomposition methods and low-rank tensor decomposition methods.

The low-rank matrix decomposition based methods have shown remarkable performances. Typically, these methods first unfold the HSI along the spectral dimension into a 2-D matrix. A denoising technique is then applied to the 2-D matrix. The low-rank matrix decomposition based methods are based on two facts, i.e., low-rank matrix decomposition can encode the high spectral correlation of HSIs well, and impulse noise, stripe noise, and dead lines all show a sparse

This research was supported by the National Natural Science Foundation of China (42001371), and the Open Fund of Hubei Luojia Laboratory (220100041). *(Corresponding author: Huanfeng Shen)*

Dongyi. Li and Xiaobin. Guan are with the School of Resource and Environmental Sciences, Hubei Luojia Laboratory, Wuhan University, Wuhan 430079, China (e-mail: DongyiLi@whu.edu.cn; guanxb@whu.edu.cn).

Dong. Chu is with Key Laboratory of Earth Surface Processes and Regional Response in the Yangtze Huaihe River Basin, Anhui Province, School of Geography and Tourism, Anhui Normal University, 241002, China(e-mail:chudong@ahnu.edu.cn).

Wei He is with the State Key Laboratory of Information Engineering in Surveying, Mapping and Remote Sensing, Wuhan University, Wuhan 430072, China.(e-mail:weihe1990@whu.edu.cn)

Huanfeng Shen is with the School of Resource and Environmental Sciences and the Collaborative Innovation Center of Geospatial Technology, Wuhan University, Wuhan 430079, China (e-mail: shenhf@whu.edu.cn).

characteristic. The low-rank matrix recovery (LRMR) method [16] lexicographically rearranges the 3-D HSI into 2-D Casorati matrices whose columns are made up of vectorized bands. It then approximates the low-rank matrix decomposition by minimizing the nuclear norm of the Casorati matrix. To improve the ability to remove non-i.i.d. noise and enhance the performance of low-rank matrix decomposition, the noise-adjusted iterative low-rank matrix approximation (NAILRMA) method [17] and the weighted Schatten p-norm (WSNLRMA) method [18] have been proposed as extensions of LRMR.

However, the low-rank matrix decomposition methods exploit the low-rank characteristic from the perspective of the spectrum, which results in insufficient utilization of the spatial information and suboptimal denoising results in the presence of severe noise. To alleviate this problem, several effective spatial prior regularization terms have been introduced into low-rank matrix decomposition models. For example, the low-rank total variation (LRTV) method [19] introduces total variation (TV) regularization into the LRMR model so that the TV regularization can efficiently preserve the spatial piecewise smoothness; the spatial-spectral TV (SSTV) regularization term extends the TV constraint to the spectrum and has achieved good results when combined with low-rank matrix decomposition [20], [21]; and the non-local self-similarity regularization (NSS) term, which exploits the similarity between non-local patches and the similarity within patches simultaneously by applying low-rank regularization for the clustering of similar patches, has also been widely combined with a low-rank matrix decomposition model, such as GLF [22] or SSLR [23]. However, all these matrix decomposition based methods need to sort the 3-D data lexicographically into 2-D data, which inevitably results in spatial structural detail information loss.

Since HSIs are 3-D tensors, tensor-based denoising methods have received increasing attention as they can effectively utilize the global spatial-spectral correlation of the HSI while efficiently restoring the spatial structure. Recently, several methods based on different tensor decomposition schemes have been proposed, such as Tucker tensor decomposition [24], tensor ring decomposition [25], tensor-singular value decomposition (t-SVD) [26], and rank-1 tensor decomposition [27]. Satisfactory denoising results can be achieved with these methods in the preservation of spatial and spectral structures and the decomposition efficiency. In addition, the tensor-based methods are often combined with other priors to achieve better results in removing complex noise. The low-rank tensor decomposition-total variation (LRTDTV) method [28] combines tensor decomposition and the TV technique by utilizing the low-rank characteristic of spatial-spectral data and the local smoothness, and has demonstrated a pleasing performance. The tensor-correlated total variation [29] method builds a unique regularization term, which essentially encodes both the global low-rank and local smoothness priors of the HSI simultaneously by applying tensor decomposition in the gradient domain. The weighted group sparsity-regularized low-rank tensor decomposition (LRTDGS) method [30] introduces weighted group sparsity regularization combined with low-rank Tucker decomposition, which can better capture the shared sparse pattern of the different images for the different bands in both spatial dimensions.

Although significant successes have been achieved with the recent HSI denoising methods, there are still some challenges that need to be addressed. Firstly, the current studies have mainly focused on exploiting the spatial-spectral correlation of the image components using various methods, but less attention has been paid to the stripe noise structure, which is often modeled as sparse noise. In fact, stripe noise exhibits high correlation in the spatial and spectral dimensions [31-35], and it spans most of the image and is no longer sparse, meaning that the current low-rank based methods cannot completely separate stripe noise from the clean HSI. The double low-rank (DLR) matrix decomposition method [3] models the stripe noise separately and encodes the spatial correlation of stripe noise by extending the LRMR method with per-band low-rank matrix decomposition regularization, obtaining great results in denoising and destriping. However, the DLR method is a matrix decomposition based method, which neglects the correlation among the stripe noise in different bands. Therefore, it is necessary to fully depict the spatial-spectral correlation of stripe noise when designing an HSI denoising model.

Secondly, most of the current HSI denoising methods based on local smoothness priors use the TV regularization based $\ell_1$-norm [36]–[39], which cannot faithfully depict the distributional structure of gradient data and usually results in the heavy-tailed phenomenon. To address this issue, the $\ell_p$ norm has been used to replace the $\ell_1$ norm to encode the spatial-spectral local smoothness of HSIs in recent research, in methods such as HyNLRMR [40] and AHTV [41]. The probability density function (PDF) of the $\ell_p$ norm is the hyper-Laplacian distribution, which can fit image gradient data well and avoid the heavy-tailed phenomenon. However, the shape of the hyper-Laplacian distribution is mostly controlled by parameter $p$, with noise remaining in high $p$ value conditions and over-smoothing in small $p$ values. Since the distribution of the gradient across distinct HSIs is different, the optimal $p$ value for hyper-Laplacian PDFs should also be different. Thus, obtaining the optimal $p$ value when applying the $\ell_p$ norm for HSI denoising is a is a key issue, and it would represent great progress if we could design a regularization term that can adaptively adjust parameter $p$ based on the characteristics of each HSI.

In this paper, to alleviate these problems, we propose an adaptive hyper-Laplacian regularized low-rank tensor decomposition (LRTDAHL) method, which can remove several types of noise, particularly stripe noise, and maintain the spatial structure well at the same time. The main contributions of this work are as follows:

1) We integrate tensor low-rank modeling of both the HSI and stripe noise into a unified model, and introduce tensor Tucker decomposition to encode the high spatial and spectral correlation in stripe noise, accurately separating it from the observed HSI.
2) We introduce an adaptive hyper-Laplacian spatial-spectral regularization (AHSSTV) term to encode the

structural sparsity in the spatial and spectral dimensions of the HSI. Unlike the previous hyper-Laplacian regularization terms, AHSSTV can adaptively estimate the hyper-Laplacian parameter of the clean gradient from the observed HSI, thereby recovering the local smoothness of the HSI more finely.

## II. THE LRTDAHL METHOD FOR HSI MIXED DENOISING

### A. Low-Rank Tensor Decomposition of Stripes and HSI

As is well known, HSIs are located in a low-dimensional space, with high redundancy in the spectral dimension. Its redundancy can be effectively represented using a low-rank tensor framework. For stripe noise in HSIs, this also has a clear low-rank feature. In order to further study the characteristics of stripe noise, we used the stripe generation method proposed in [31] to simulate a set of stripe noise so that we could study the correlation of stripes from a quantitative perspective. Fig. 1 shows the simulated stripe noise, the unfolding matrices of the spatial and spectral dimensions, and the corresponding singular value curves. The decay of the singular value curves of the unfolding matrices proves the strong correlation of stripe noise in the two spatial dimensions and the relatively weaker correlation in the spectral dimension. Based on the prior knowledge and mathematical research into the spatial-spectral correlation of tensor data, we introduce low-rank Tucker decomposition to aggregate the decomposed matrices obtained as described above:

$$\mathcal{B} = \mathcal{U} \times_1 G_1 \times_2 G_2 \times_3 G_3 \quad (1)$$

where $\mathcal{U}$ is the abundance tensor, $G_i$ is the endmember matrix. Compared to modeling stripe noise with per-band low-rank matrix decomposition, Tucker decomposition can effectively encode the low-rank information in both spatial dimensions, resulting in more accurate extraction of stripe noise from HSIs that contain noise [34].

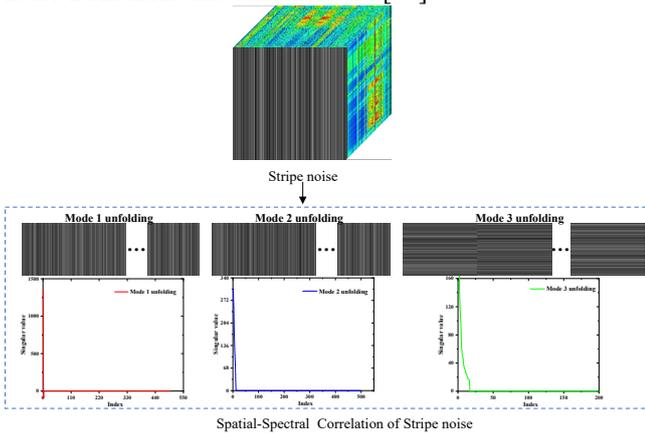

Fig. 1. The spatial low-rank priors of stripes.

### B. HSI Observation Model

The traditional processing methods typically decompose the obtained HSI with mixed noise $\mathcal{Y}$ into clean image $\mathcal{X}$, Gaussian noise $\mathcal{N}$, and sparse noise $\mathcal{S}$, with the stripe noise categorized as sparse noise $\mathcal{S}$. However, previous studies [35] have indicated that the generation and actual distribution of stripe noise are different from those of sparse noise, and they separated the stripe noise from $\mathcal{X}$ and modeled it separately as matrix $\mathcal{B}$. Based on the analysis results presented in Section II-A, it is evident that stripe noise exhibits a certain level of correlation in both the spatial and spectral dimensions. Furthermore, compared to matrices, tensors can more precisely capture the low-rank structure of stripe noise. Therefore, we propose to model the stripe noise $\mathcal{B}$ using tensors, extending the previous observation model into a double tensor model. The HSI observation model can then be rewritten as follows:

$$\mathcal{Y} = \mathcal{X} + \mathcal{S} + \mathcal{N} + \mathcal{B} \quad (2)$$

where $\mathcal{Y} \in \mathbb{R}^{h \times w \times p}$ denotes the observed HSI data; $h$ and $w$ denote the height and width of the $i$-th band, respectively; and $p$ denotes the number of bands.

### C. Adaptive Hyper-Laplacian Total Variation Prior

Based on the fact that image gradients present a sparse distribution, TV prior regularization and its extensions have been widely used in HSI denoising methods [19], among which SSTV takes into account both the spatial and spectral gradient information and demonstrates a good performance. Specifically, SSTV can be written as:

$$\|\mathcal{X}\|_{sstv} = \sum_{i,j,k} w_h |\mathcal{X}_{i,j,k} - \mathcal{X}_{i-1,j,k}| + w_w |\mathcal{X}_{i,j,k} - \mathcal{X}_{i,j-1,k}| + w_p |\mathcal{X}_{i,j,k} - \mathcal{X}_{i,j,k-1}| \quad (3)$$

where $(w_h, w_w, w_b)$ represent the weight along the three modes of $\mathcal{X}$ that controls its gradient sparsity, which are usually set to (1, 1, 0.5) [20]. SSTV regularization is implemented through the $\ell_1$ norm, which results in the data distribution tending toward a Laplacian distribution, whose PDF is $p(x;\lambda) \propto e^{\lambda|x|}$. However, recent research has shown that the statistical distributions of both the spatial and spectral gradients are more consistent with a super-Laplacian distribution, with PDF $p(x;p) \propto ke^{-k|x|^p}$.

Fig. 2 shows the histograms of the gradient in three directions (spatial and spectral) and the fitting of the three different probability density curves: Gaussian, Laplacian, and super-Laplacian. As shown in Fig. 2, the super-Laplacian distribution can more effectively fit the distribution characteristics of the three gradient directions of the HSI. The super-Laplacian PDFs fitted by different HSIs and different dimensions of the same HSI's gradient distribution characteristics will be different. Clearly, the closer the super-Laplacian PDF fit is to the real gradient distribution characteristic, the better the denoising effect.

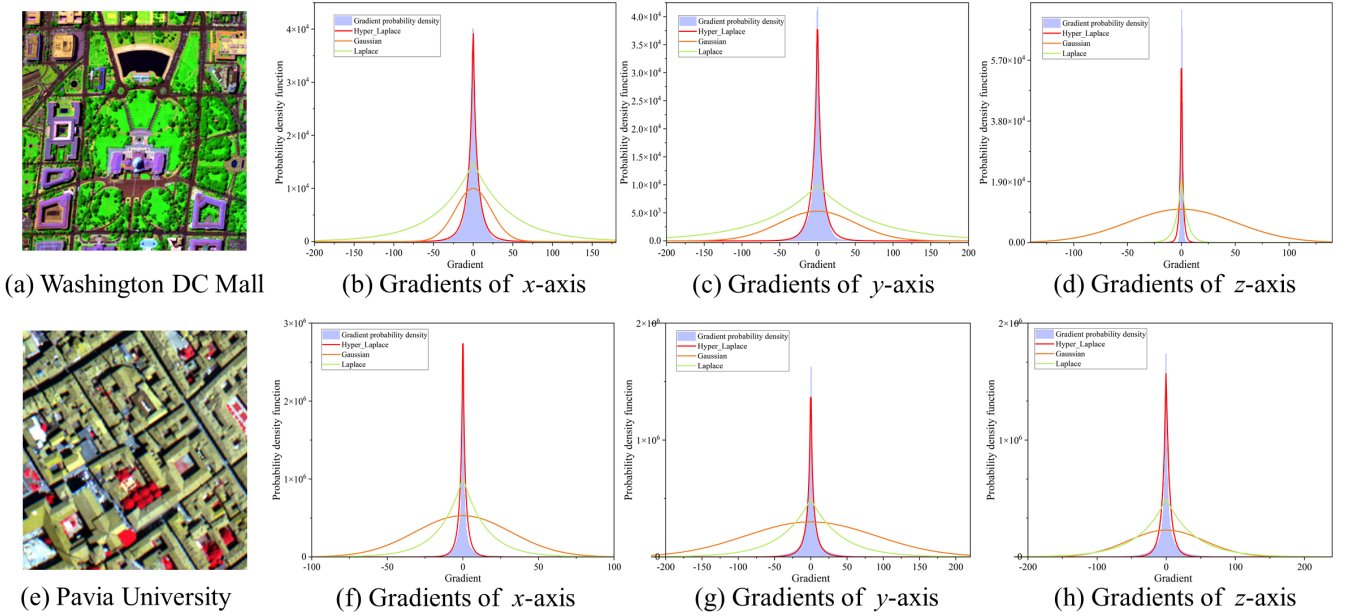

Fig. 2. Spatial-spectral gradient distributions and probability density functions of the HSIs. (a) and (e) are the false-color images for the HSIs. (b)–(d) and (f)–(h) are the HSI histograms of gradients along the spatial and spectral dimensions and the fitted probability density curves, respectivel

This means that, when designing a super-Laplacian regularizer, it is necessary to adaptively estimate the super-Laplacian parameters based on the gradient characteristics of the different dimensions of different images. By maximizing the likelihood function of the super-Laplacian distribution, it is possible to derive the adaptive spectral-spatial total variation (AHSSTV) term as follows:

$$\|\mathcal{X}\|_{AHSSTV} = w_h \|\mathcal{D}_x \mathcal{X}\|_{p_x}^{p_x} + w_w \|\mathcal{D}_y \mathcal{X}\|_{p_y}^{p_y} + w_b \|\mathcal{D}_z \mathcal{X}\|_{p_z}^{p_z}, \quad (6)$$

where $\mathcal{D}_x$, $\mathcal{D}_y$, and $\mathcal{D}_z$ are the 3-D difference operators. $p_x$, $p_y$, and $p_z$ are the 3-D hyper-Laplacian parameters. $p \in (0,1)$, and when $p = 1$, AHSSTV degenerates to SSTV. We propose a method for adaptively estimating $p$ based on the relevant research [42], [43].

1) For the observed HSI $\mathcal{Y}$, the clean image $\mathcal{X}$, and the noise $\mathcal{N}$, by performing differential operations in the spatial and spectral dimensions, three difference variables are constructed, $Y = [Y_i | i = 1,2,3] \in \mathbb{R}^{n_b \times p \times 3}$ $Y_i = \mathcal{D}_i(\mathcal{Y})$, $X = [X_i | i = 1,2,3] \in \mathbb{R}^{n_b \times p \times 3}$, $X_i = \mathcal{D}_i(\mathcal{X})$, $N = [N_i | i = 1,2,3] \in \mathbb{R}^{n_b \times p \times 3}$, $N_i = \mathcal{D}_i(\mathcal{N})$, where $\mathcal{D}_i(\cdot)$ is the difference operator in the $i$-th dimension and $n_b$ is the number of pixels in each spectral band. Based on the relevant research on noise estimation, a noise estimator is applied to Y to obtain the variance of the noise $\sigma = [\sigma_1, \sigma_2, \sigma_3]$, where $\sigma_i = esn(Y_i)$ and $esn(\cdot)$ is the noise variance estimator.

2) Since the clean image $\mathcal{Y}$ and noise $\mathcal{N}$ are independent of each other, it can be assumed that $Y_i$ and $N_i$ are mutually independent. At the same time, it can be assumed that $Y_i$ and $N_i$ follow a Laplacian distribution $f_{X_i} = k_i \exp(-k_i |X_i|^{p_i})$ and Gaussian distribution $f_{N_i} = (1/2\sqrt{\pi}\sigma_i)\exp(-N_i/4\sigma_i^2)$, respectively, and their PDF $f_{Y_i}$ can be written as:

$$f_{Y_i} = f_{X_i} \times f_{N_i} \quad (7)$$

The PDF of $Y_i$ is:

$$f_{Y_i}(Y_i = a) = \int_t f_{X_i}(X_i = t) \times f_{N_i}(N_i = a - t)dt \quad (8)$$

Furthermore, due to the numerical discreteness of the image gradient, the probability density $f$ with a normalized discrete histogram $h$ can be approximated. The above equation can then be transformed into:

$$h_{Y_i} = h_{X_i} \otimes h_{N_i} \quad (9)$$

where $\otimes$ represents the convolution operator, and $h_{Y_i}$, $h_{N_i}$ can be calculated based on the observed HSI and noise variance. Furthermore, since $X_i$ follows a Laplacian distribution, it is possible to obtain $h_{X_i} = h_f(k_i, p_i)$.

3) The goal is to estimate the optimal Laplacian distribution parameters $P = [p_i | i = 1,2,3]$ for the spatial-spectral gradient image in the three dimensions, which can be written as:

$$\{k_i, p_i\} = \arg\min_{k_i, p_i} \| h_{Y_i} - h_f(k_i, p_i) \otimes h_{N_i} \|_F^2 \quad (10)$$

where $k_i$ and $p_i$ are the parameters of the Laplacian distribution, but we only need parameter $p_i$ in the derived Laplacian regularizer. Given that the gradient data of the image have a narrow range for both $k_i$ and $p_i$, the Nelder-Mead algorithm [43] is used to search for these parameters and determine the optimal values. The impact of adaptive parameter $p_i$ estimation on the denoising performance is further analyzed in Section III.

*D. Optimization Model*

As stated above, utilizing prior knowledge of the image to design an effective regularizer is the key to denoising HSIs.

Based on the discussions in Sections III-A and III-C, we propose the *adaptive hyper-Laplacian regularized low-rank tensor decomposition* (LRTDAHL) model:

$$\arg\min_{\mathcal{X},\mathcal{B},\mathcal{S}} \frac{1}{2}\|\mathcal{Y}-\mathcal{X}-\mathcal{S}-\mathcal{B}\|_F^2 + \lambda_1\|\mathcal{X}\|_{AHSSTV} + \lambda_2\|\mathcal{S}\|_1 + \\ \|\mathcal{X}\|_* + \|\mathcal{B}\|_* \quad (11)\\ s.t. \mathcal{X} = \mathcal{C}\times_1 V_1 \times_2 V_2 \times_3 V_3, V_i^T V_i = I(i=1,2,3), \\ \mathcal{B} = \mathcal{U}\times_1 G_1 \times_2 G_2 \times_3 G_3, G_i^T G_i = I(i=1,2,3)$$

where $\lambda_1$ and $\lambda_2$ are the regularization parameters that control the adaptive hyper-Laplacian term and the $\ell_1$ regularization term, respectively; $\|\cdot\|_*$ represents the low-rank tensor regularization term; $\|\cdot\|_{AHSSTV}$ represents the adaptive hyper-Laplacian regularization term; and $\|\cdot\|_1$ represents the $\ell_1$ regularization term, which is used to remove the sparse noise. It is worth noting that the LRTDAHL model considers the prior information of the clean image, Gaussian noise, sparse noise, and stripe noise, and is expected to have the ability to remove multiple types of mixed noise. Specifically, we consider the prior information of the spatial-spectral low rank of the clean image and the spatial low rank of stripe noise, and introduce Tucker tensor low-rank decomposition to model them, which can fully encode the spatial-spectral low-rank characteristics of the clean HSI and the spatial low-rank characteristic of stripe noise. In addition, the AHSSTV regularization term can adaptively estimate the parameters required for the hyper-Laplacian regularization, which is better at preserving detail information and suppressing Gaussian noise and sparse noise, compared to $\ell_1$.

### E. Optimization Procedure

As the proposed model is clearly a non-convex optimization problem, there are various methods for solving non-convex optimization problems. Based on the augmented Lagrange multiplier (ALM) optimization method (11), some auxiliary variables are first introduced to reformulate (11) as:

$$\arg\min_{\mathcal{X},\mathcal{B},\mathcal{S}} \frac{1}{2}\|\mathcal{Y}-\mathcal{Z}-\mathcal{S}-\mathcal{B}\|_F^2 + \lambda_1\|\mathcal{F}\|_{AHSSTV} + \lambda_2\|\mathcal{S}\|_1 + \\ \|\mathcal{X}\|_* + \|\mathcal{B}\|_* + <\Lambda_1, \mathcal{X}-\mathcal{Z}> + <\Lambda_1, \mathcal{D}_w(\mathcal{Z})-\mathcal{F}> \\ \frac{\beta}{2}(\|\mathcal{X}-\mathcal{Z}\|_F^2 + \|\mathcal{D}_w(\mathcal{Z})-\mathcal{F}\|_F^2) \quad (12)\\ s.t. \mathcal{Z}=\mathcal{X}, \mathcal{D}_w(\mathcal{Z})=\mathcal{F}, \mathcal{X} = \mathcal{C}\times_1 V_1 \times_2 V_2 \times_3 V_3, \\ V_i^T V_i = I(i=1,2,3), \mathcal{B} = \mathcal{U}\times_1 G_1 \times_2 G_2 \times_3 G_3, \\ G_i^T G_i = I(i=1,2,3)$$

where $\Lambda_1$ and $\Lambda_2$ are the Lagrange multipliers, $\beta$ is a positive penalty parameter, and $\mathcal{D}_w(\cdot) = [w_h \mathcal{D}_h(\cdot); w_w \mathcal{D}_w(\cdot); w_z \mathcal{D}_z(\cdot);]$ is a weighted difference operator. One solution to this problem is to iteratively optimize the augmented Lagrangian function on one variable while fixing the others, which is known as the alternating direction method of multipliers (ADMM) algorithm. Specifically, in the $k+1$-th iteration, the variables are updated as follows:

1) *Update* $\mathcal{X}$ :

$$\arg\min_{\mathcal{X}} \|\mathcal{X}\|_* + \frac{\beta}{2}\|\mathcal{X} - \frac{\beta\mathcal{Z}^k - \Lambda_1^k}{\beta}\|_F^2 \quad (13)$$

Tucker tensor decomposition is introduced, and (13) can be transformed into:

$$\arg\min_{\mathcal{X}} \|\mathcal{X}\|_* + \frac{\beta}{2}\|\mathcal{C}\times_1 V_1 \times_2 V_2 \times_3 V_3 - \frac{\beta\mathcal{Z}^k - \Lambda_1^k}{\beta}\|_F^2 \quad (14)$$

By using the higher-order orthogonal iteration (HOOI) algorithm [28], $\mathcal{C}^{k+1}$ and $V_i^{k+1}$ can be obtained, and $\mathcal{X}$ can be updated as follows:

$$\mathcal{X}^{k+1} = \mathcal{C}^{k+1}\times_1 V_1^{k+1} \times_2 V_2^{k+1} \times_3 V_3^{k+1} \quad (15)$$

2) *Update* $\mathcal{Z}$ :

$$\arg\min_{\mathcal{Z}} \frac{1}{2}\|\mathcal{Y}-\mathcal{Z}-\mathcal{S}^k-\mathcal{B}^k\|_F^2 + \frac{\beta}{2}(\|\mathcal{X}^k-\mathcal{Z}\|_F^2 + \|\mathcal{D}_w(\mathcal{Z})-\mathcal{F}^k\|_F^2) \\ + <\Lambda_1^k, \mathcal{X}^k-\mathcal{Z}> + <\Lambda_2^k, \mathcal{D}_w(\mathcal{Z})-\mathcal{F}^k> \quad (16)$$

This problem can be viewed as a least-squares problem, and through derivation, the following can be obtained:

$$\arg\min_{\mathcal{Z}} \frac{\beta+1}{2}\|\mathcal{Z} - \frac{\mathcal{Y}-\mathcal{B}^k-\mathcal{S}^k+\Lambda_1^{k+1}+\beta\mathcal{X}^{k+1}}{\beta+1}\|_F^2 \\ + \frac{\beta}{2}\|\mathcal{D}_w(\mathcal{Z}) - \frac{\beta F^k - \Lambda_2^k}{\beta}\|_F^2 \quad (17)$$

The items in (17) are expanded and similar terms merged to obtain:

$$(I + \beta + \beta\mathcal{D}_w^T\mathcal{D}_w)\mathcal{Z} = (\mathcal{Y} - \mathcal{B}^k - \mathcal{S}^k + \Lambda_1^{k+1} + \beta\mathcal{X}^{k+1}) \\ + \beta D_w^T(\mathcal{F}^k - \Lambda_2^k/\beta) \quad (18)$$

where $\mathcal{D}_w^T$ is the companion matrix of $\mathcal{D}_w$, and $\mathcal{D}_w^T\mathcal{D}_w$ has a block matrix structure. To speed up the calculation, 3-D fast Fourier transform (FFT) can be used to solve the problem as follows:

$$\begin{cases} H_\mathcal{Z} = (\mathcal{Y}-\mathcal{B}^k-\mathcal{S}^k+\Lambda_1^{k+1}+\beta\mathcal{X}^{k+1}) + \beta D_w^T(\mathcal{F}^k - \Lambda_2^k/\beta) \\ T_\mathcal{Z} = |fftn(\mathcal{D}_x)|^2 + |fftn(\mathcal{D}_y)|^2 + |fftn(\mathcal{D}_z)|^2 \\ Z^{k+1} = ifftn(\frac{fftn(H_\mathcal{Z})}{\beta+1+\beta T_\mathcal{Z}}) \end{cases} \quad (19)$$

where $fftn(\cdot)$ and $ifftn(\cdot)$ denote FFT and inverse FFT, respectively.

3) *Update* $\mathcal{F}$ :

$$\arg\min_{\mathcal{F}} \lambda_1\|\mathcal{F}\|_{AHSSTV} + \frac{\beta}{2}\|\mathcal{F} - \frac{\beta\mathcal{D}_w(\mathcal{Z}^{k+1})+\Lambda_2^k}{\beta}\|_F^2 \quad (20)$$

The following spatial-spectral difference operator is introduced:

$$L_\theta(\gamma, p_x, p_y, p_z) = [\|\gamma_{dx}\|_{p_x}^{p_x}; \|\gamma_{dy}\|_{p_y}^{p_y}; \|\gamma_{dz}\|_{p_z}^{p_z}] \quad (21)$$

where $\gamma_{dx}$, $\gamma_{dy}$, and $\gamma_{dz}$ represent the decoding operators of the three dimensions of the spatial-spectral difference operator. $\|\cdot\|_p^p$ (0<p<1) is a non-convex norm minimization problem, and the generalized iterative shrinkage algorithm [42] can lead to an accurate and convergent solution.

Therefore, the following can be obtained:

$$\mathcal{F}^{k+1} = L_{\frac{\lambda_1}{\beta}}(\frac{\beta\mathcal{D}_w(\mathcal{Z}^{k+1})+\Lambda_2^k}{\beta}, p_h, p_w, p_p) \quad (22)$$

where $p_h$, $p_w$ and $p_p$ are the 3-D gradient hyper-Laplacian parameters estimated through (8) for the HSI.

4) *Update* $\mathcal{B}$ :

$$\arg\min_{\mathcal{B}} \frac{\beta}{2}\|\mathcal{Y}-\mathcal{Z}^{k+1}-\mathcal{S}^k-\mathcal{B}\|_F^2 + \|\mathcal{B}\|_* \quad (23)$$

By introducing Tucker tensor decomposition, (23) can be transformed into:

$$\arg\min_{\mathcal{U},G_i} \frac{\beta}{2} \| \mathcal{U} \times_1 G_1 \times_2 G_2 \times_3 G_3 - (\mathcal{Y}-\mathcal{Z}^{k+1}-\mathcal{S}^k) \|_F^2 + \| \mathcal{B} \|_* \quad (24)$$

By using the HOOI algorithm, $\mathcal{U}^{k+1}$ and $G_i^{k+1}$ can be obtained, and $\mathcal{B}$ can be updated as follows:

$$\mathcal{B}^{k+1} = \mathcal{U}^{k+1} \times_1 G_1^{k+1} \times_2 G_2^{k+1} \times_3 G_3^{k+1} \quad (25)$$

5) *Update* $\mathcal{S}$:

$$\arg\min_{\mathcal{X},\mathcal{B},\mathcal{S}} \frac{1}{2} \| \mathcal{Y}-\mathcal{Z}^{k+1}-\mathcal{B}^{k+1}-\mathcal{S} \|_F^2 + \lambda_2 \| \mathcal{S} \|_1 \quad (26)$$

The following soft threshold operator is then introduced:

$$\Re_\Delta(\mathcal{X}) = \begin{cases} \mathcal{X} - \Delta, & \mathcal{X} > \Delta \\ \mathcal{X} + \Delta, & \mathcal{X} < \Delta \\ 0, & \mathcal{X} = \Delta \end{cases} \quad (27)$$

The closed-form solution of the problem (26) can be obtained by (27), i.e.,

$$\mathcal{S}^{k+1} = \Re_\Delta(\mathcal{Y}-\mathcal{Z}^{k+1}-\mathcal{B}^{k+1}) \quad (28)$$

6) *Update* the Lagrangian:

$$\begin{cases} \Lambda_1^{k+1} = \Lambda_1^k + \beta(\mathcal{X}^{k+1} - \mathcal{Z}^{k+1}) \\ \Lambda_2^{k+1} = \Lambda_2^k + \beta(D_w(\mathcal{Z}^{k+1}) - \mathcal{F}^{k+1}) \end{cases} \quad (29)$$

Combining the sub-problems solved above, the solution to the optimization process can be summarized as shown in Algorithm 1.

---

**Algorithm 1:** Optimization procedure for the proposed model.

1: **Input:** Noisy HSI $\mathcal{Y}$, Tucker desired rank [$r_1,r_2,r_3$] and [$r_a,r_b,r_c$], parameters $\varepsilon, \beta, \lambda_1, \lambda_2, k_{max}$
2: **Initialize:** $\mathcal{X} = \mathcal{B} = \mathcal{S} = \mathcal{F} = \mathcal{Z} = 0, \Lambda_1 = \Lambda_2 = 0, \mu_{max} = 10^6, k=0$
3: **Estimate p:** Estimate $p_h, p_w, p_p$ via (8)
4: **While** stopping criterion is not satisfied **do**:
5: Update $\mathcal{X}, \mathcal{C}, V_i$ via (14)–(15)
6: Update $\mathcal{Z}$ via (19)
7: Update $\mathcal{F}$ via (22)
8: Update $\mathcal{B}$ via (24)–(25)
9: Update $\mathcal{S}$ via (28)
10: Update the Lagrangian multipliers $\Lambda_1, \Lambda_2$ via (29)
11: Check the convergence condition $\frac{\| \mathcal{X}^{k+1} - \mathcal{X}^k \|_F^2}{\| \mathcal{X}^k \|_F^2} \leq \varepsilon$
12: **Output:** The restored HSI

---

## III. EXPERIMENTAL RESULTS AND DISCUSSION

In this section, we describe how we tested the effectiveness of the proposed LRTDAHL model based on simulated and real-data experiments, and compared it with eight other state-of-the-art HSI denoising methods: total-variation-regularized low-rank matrix factorization (LRTV) [19], total-variation-regularized low-rank tensor decomposition (LRTDTV) [28], three-directional log-based tensor nuclear norm (3DlogTNN) [44], a fast and parameter-free hyperspectral image mixed noise removal method (FastHyMix) [45], factor group sparsity-regularized non-convex low-rank approximation (FGSLR) [46], the non-local transform-domain filter (BM4D) [47], non-local meets global (NGmeet) [48], non-i.i.d. mixture of Gaussians (NMoG) [49], double low-rank matrix decomposition (DLR) [3], and fast Graph Laplacian Regularization (FGLR) [4]. For all the comparison methods, the parameter settings were adjusted according to the suggestions in the literature. The parameter selection for the proposed LRTDAHL model is discussed in the Section III.C.

### A. Simulated Data Experiments

*1) Experimental Settings:* In order to evaluate the noise removal performance of the proposed LRTDAHL model, we conducted simulated experiments and compared the visual and quantitative performances. Three representative clean HSI datasets were used in the experiments to simulate noisy HSIs, namely, the Washington DC Mall dataset, ROSIS Pavia City Center Dataset, and the Indian Pines dataset. There can be several different types of noise in real HSIs, including Gaussian noise, impulse noise, dead lines, and stripe noise. In order to simulate these scenarios as closely as possible, we compared six different simulated noise situations. Before the noise simulation, the gray values of each band were normalized to [0,1].

In order to simulate noise in the HSIs, we designed the following six cases.

TABLE I
SIMULATED NOISE CASES

| Simulated case | Added noise | | |
|---|---|---|---|
| | $\sigma^2$ of Gaussian noise | Percentages of impulse noise | Intensity of stripes |
| Case1 | 0–0.2 | 0–0.2 | 0 |
| Case2 | 0.1 | 0.2 | 40–50% |
| Case3 | 0–0.2 | 0–0.2 | 60–70% |
| Case4 | 0–0.2 | 0–0.2 | 40% periodic |
| Case5 | 0–0.2 | 0–0.2 | 40% mixed |
| Case6 | 0–0.2 | 0–0.2 | Wide vertical stripes |

*Case1 (Gaussian Noise + impulse noise):* Gaussian noise, dead lines, and impulse noise were added as illustrated in Table I, to assess the performance of the proposed method and the comparison methods in the absence of stripes.

*Case2 (Gaussian noise + impulse noise + dead lines + stripes):* Gaussian noise, dead lines, 40–50% stripes, and impulse noise were added as illustrated in Table I, to assess the performance of the proposed method and the comparison methods under the condition of relatively low-intensity stripes.

*Case3 (Gaussian Noise + impulse noise + dead lines + stripes):* Gaussian noise, dead lines, 60–70% stripes, and impulse noise were added as illustrated in Table I, to assess the performance of the proposed method and the comparison methods under the condition of relatively high-intensity stripes.

*Case4 (Gaussian Noise + impulse noise + dead lines + periodic stripes)*: Gaussian noise, dead lines, 40% periodic stripes, and impulse noise were added as illustrated in Table I, to simulate the performance of the proposed method and the comparison methods in the presence of periodic stripes.

*Case5 (Gaussian Noise + impulse noise + dead lines + mixed stripes):* Gaussian noise, dead lines, 40% mixed stripes, and impulse noise were added as illustrated in Table I, to simulate the performance of the proposed method and the comparison methods in the presence of mixed stripes.

*Case6 (Gaussian Noise + impulse noise + dead lines + wide vertical stripes):* Gaussian noise, dead pixels, wide vertical stripes, and impulse noise were added as illustrated in Table I, to simulate the performance of the proposed method and the comparison methods in the presence of wide vertical stripes. *2) Quantitative Comparison:* The quantitative evaluation results of the various methods on the simulated experimental datasets under the different noise conditions are presented in Table II, Table III, and Table IV.

TABLE II
QUANTITATIVE ASSESSMENT RESULTS OF ALL THE METHODS ON THE PAVIA CITY CENTER DATASET

| Case | Evaluation index | Noisy | BM4D | NMoG | NGmeet | LRTV | LRTDTV | 3DlogTNN | FGLR | DLR | FGSLR | FastHyMix | LRTDAHL |
|---|---|---|---|---|---|---|---|---|---|---|---|---|---|
| Case1 | MPSNR | 14.72 | 27.15 | 38.92 | 39.21 | 36.95 | 37.04 | 38.99 | 37.85 | 36.91 | **40.14** | 39.33 | 39.04 |
| | MSSIM | 0.261 | 0.785 | 0.979 | 0.988 | 0.968 | 0.964 | 0.981 | 0.961 | 0.969 | **0.994** | 0.992 | 0.980 |
| | MSAM | 0.723 | 0.201 | **0.071** | 0.088 | 0.097 | 0.095 | 0.098 | 0.091 | 0.115 | 0.079 | 0.087 | 0.081 |
| Case2 | MPSNR | 10.05 | 20.11 | 33.57 | 32.88 | 30.11 | 29.90 | 30.45 | 34.91 | 35.58 | 35.47 | 30.98 | **36.17** |
| | MSSIM | 0.088 | 0.741 | 0.931 | 0.922 | 0.907 | 0.896 | 0.884 | 0.941 | 0.956 | 0.957 | 0.915 | **0.966** |
| | MSAM | 0.853 | 0.369 | 0.139 | 0.177 | 0.239 | 0.244 | 0.256 | 0.116 | 0.141 | 0.131 | 0.199 | **0.104** |
| Case3 | MPSNR | 11.78 | 19.69 | 31.41 | 30.28 | 25.31 | 26.36 | 25.64 | 31.67 | 31.73 | 31.37 | 28.37 | **32.79** |
| | MSSIM | 0.171 | 0.745 | 0.943 | 0.906 | 0.891 | 0.887 | 0.869 | 0.931 | 0.943 | 0.944 | 0.867 | **0.961** |
| | MSAM | 0.875 | 0.479 | 0.195 | **0.118** | 0.345 | 0.291 | 0.245 | 0.148 | 0.232 | 0.144 | 0.187 | 0.122 |
| Case4 | MPSNR | 12.24 | 20.11 | 27.92 | 28.12 | 23.73 | 23.81 | 23.34 | 28.75 | 30.11 | 27.29 | 25.32 | **31.92** |
| | MSSIM | 0.166 | 0.704 | 0.869 | 0.866 | 0.875 | 0.835 | 0.834 | 0.895 | 0.902 | 0.861 | 0.849 | **0.924** |
| | MSAM | 0.782 | 0.382 | 0.383 | 0.441 | 0.363 | 0.352 | 0.427 | 0.284 | 0.201 | 0.399 | 0.378 | **0.195** |
| Case5 | MPSNR | 12.95 | 22.67 | 33.31 | 31.46 | 25.95 | 26.87 | 27.70 | 33.01 | 32.74 | 32.43 | 29.58 | **34.98** |
| | MSSIM | 0.182 | 0.768 | 0.904 | 0.882 | 0.821 | 0.815 | 0.831 | 0.907 | 0.896 | 0.894 | 0.870 | **0.912** |
| | MSAM | 0.842 | 0.396 | 0.214 | 0.222 | 0.311 | 0.301 | 0.275 | 0.201 | 0.196 | 0.214 | 0.353 | **0.184** |
| Case6 | MPSNR | 14.22 | 23.13 | 30.94 | 32.83 | 27.03 | 27.92 | 28.11 | 31.72 | 32.79 | 31.62 | 29.37 | **33.61** |
| | MSSIM | 0.242 | 0.699 | 0.908 | 0.915 | 0.868 | 0.901 | 0.906 | 0.916 | 0.911 | **0.938** | 0.920 | 0.929 |
| | MSAM | 0.712 | 0.336 | 0.174 | 0.155 | 0.244 | 0.233 | 0.211 | 0.199 | 0.176 | 0.149 | 0.153 | **0.146** |

TABLE III
QUANTITATIVE ASSESSMENT RESULTS OF ALL THE METHODS ON THE INDIAN PINES DATASET

| Case | Evaluation index | Noisy | BM4D | NMoG | NGmeet | LRTV | LRTDTV | 3DlogTNN | FGLR | DLR | FGSLR | FastHyMix | LRTDAHL |
|---|---|---|---|---|---|---|---|---|---|---|---|---|---|
| Case1 | MPSNR | 14.44 | 27.45 | 38.30 | 40.44 | 36.66 | 37.56 | 36.79 | 37.41 | 35.81 | **42.37** | 39.66 | 41.05 |
| | MSSIM | 0.276 | 0.813 | 0.987 | 0.983 | 0.969 | 0.975 | 0.966 | 0.961 | 0.951 | 0.986 | **0.989** | 0.982 |
| | MSAM | 0.742 | 0.257 | 0.081 | 0.085 | 0.116 | 0.124 | 0.091 | 0.140 | 0.144 | **0.079** | 0.085 | 0.087 |
| Case2 | MPSNR | 11.72 | 19.87 | 33.22 | 33.59 | 30.14 | 31.55 | 31.46 | 34.01 | 34.91 | 34.10 | 33.22 | **35.59** |
| | MSSIM | 0.167 | 0.733 | 0.951 | 0.957 | 0.902 | 0.909 | 0.911 | 0.949 | 0.954 | 0.941 | 0.921 | **0.981** |
| | MSAM | 0.759 | 0.577 | 0.161 | 0.198 | 0.277 | 0.274 | 0.256 | 0.211 | 0.221 | 0.197 | 0.241 | **0.133** |
| Case3 | MPSNR | 13.01 | 18.77 | 33.23 | 32.46 | 29.99 | 29.48 | 28.54 | 33.21 | 32.71 | 33.15 | 32.51 | **35.14** |
| | MSSIM | 0.206 | 0.723 | 0.923 | 0.911 | 0.907 | 0.912 | 0.883 | 0.941 | 0.934 | 0.935 | 0.913 | **0.962** |
| | MSAM | 0.768 | 0.466 | 0.155 | 0.187 | 0.264 | 0.245 | 0.301 | 0.223 | 0.249 | 0.184 | 0.255 | **0.139** |
| Case4 | MPSNR | 13.47 | 19.36 | 30.11 | 31.45 | 27.36 | 27.99 | 27.36 | 32.14 | 33.14 | 31.65 | 29.66 | **36.11** |
| | MSSIM | 0.241 | 0.756 | 0.931 | 0.899 | 0.887 | 0.891 | 0.871 | 0.932 | 0.951 | 0.904 | 0.876 | **0.964** |
| | MSAM | 0.788 | 0.511 | 0.186 | 0.211 | 0.268 | 0.274 | 0.301 | 0.194 | 0.188 | 0.178 | 0.289 | **0.124** |
| Case5 | MPSNR | 13.56 | 20.22 | 32.25 | 29.24 | 30.14 | 30.79 | 27.69 | 33.01 | 33.21 | 33.14 | 32.64 | **34.71** |
| | MSSIM | 0.229 | 0.744 | 0.922 | 0.891 | 0.905 | 0.910 | 0.854 | 0.959 | 0.954 | 0.955 | 0.932 | **0.969** |
| | MSAM | 0.786 | 0.532 | 0.176 | 0.187 | 0.199 | 0.201 | 0.314 | 0.184 | 0.191 | 0.198 | 0.297 | **0.165** |
| Case6 | MPSNR | 14.44 | 18.97 | 31.48 | 30.45 | 29.14 | 30.15 | 30.44 | 32.31 | 32.11 | 32.14 | 31.22 | **33.40** |
| | MSSIM | 0.271 | 0.786 | 0.931 | 0.905 | 0.896 | 0.913 | 0.915 | 0.951 | 0.941 | 0.947 | 0.939 | **0.966** |
| | MSAM | 0.741 | 0.469 | 0.201 | 0.181 | 0.251 | 0.243 | 0.248 | 0.201 | 0.223 | 0.196 | 0.214 | **0.170** |

TABLE IV
QUANTITATIVE ASSESSMENT RESULTS OF ALL THE METHODS ON THE WASHINGTON DC MALL DATASET

| Case | Evaluation index | Noisy | BM4D | NMoG | NGmeet | LRTV | LRTDTV | 3DlogTNN | FGLR | DLR | FGSLR | FastHyMix | LRTDAHL |
|---|---|---|---|---|---|---|---|---|---|---|---|---|---|
| Case1 | MPSNR | 14.12 | 29.11 | 37.88 | 40.11 | 36.11 | 36.12 | 39.12 | 37.59 | 35.57 | **40.32** | 39.96 | 39.83 |
| | MSSIM | 0.231 | 0.795 | 0.954 | 0.974 | 0.921 | 0.932 | 0.945 | 0.938 | 0.919 | 0.971 | 0.961 | 0.964 |
| | MSAM | 0.702 | 0.375 | 0.124 | 0.111 | 0.156 | 0.125 | 0.134 | 0.141 | 0.161 | **0.098** | 0.102 | 0.101 |
| Case2 | MPSNR | 11.80 | 18.09 | 31.99 | 30.54 | 25.77 | 26.99 | 29.42 | 32.14 | 31.61 | 32.93 | 29.88 | **34.41** |

|  |  |  |  |  |  |  |  |  |  |  |  |  |
|---|---|---|---|---|---|---|---|---|---|---|---|---|
|  | MSSIM | 0.109 | 0.701 | 0.901 | 0.891 | 0.859 | 0.864 | 0.897 | <u>0.922</u> | 0.191 | 0.914 | 0.892 | **0.933** |
|  | MSAM | 0.787 | 0.546 | 0.184 | 0.198 | 0.301 | 0.274 | 0.201 | <u>0.151</u> | 0.164 | 0.156 | 0.198 | **0.114** |
| Case3 | MPSNR | 13.95 | 16.53 | 30.17 | 28.77 | 22.97 | 23.47 | 24.45 | <u>31.04</u> | 30.91 | 29.84 | 29.13 | **32.41** |
|  | MSSIM | 0.181 | 0.657 | <u>0.894</u> | 0.889 | 0.844 | 0.847 | 0.849 | 0.919 | 0.901 | 0.891 | 0.887 | **0.946** |
|  | MSAM | 0.683 | 0.511 | 0.207 | 0.221 | 0.289 | 0.279 | 0.256 | <u>0.181</u> | 0.191 | 0.201 | 0.212 | **0.169** |
| Case4 | MPSNR | 14.35 | 19.79 | 29.54 | 27.66 | 24.55 | 25.69 | 26.38 | <u>30.59</u> | 30.41 | 29.47 | 28.31 | **32.18** |
|  | MSSIM | 0.208 | 0.645 | 0.899 | 0.866 | 0.841 | 0.845 | 0.867 | <u>0.911</u> | 0.902 | 0.901 | 0.876 | **0.922** |
|  | MSAM | 0.730 | 0.432 | 0.231 | 0.246 | 0.304 | 0.287 | 0.244 | 0.247 | 0.231 | 0.234 | <u>0.219</u> | **0.201** |
| Case5 | MPSNR | 13.32 | 19.84 | 32.02 | 29.55 | 26.01 | 26.36 | 29.69 | 31.48 | 31.74 | <u>32.11</u> | 30.29 | **33.62** |
|  | MSSIM | 0.182 | 0.597 | 0.907 | 0.877 | 0.870 | 0.861 | 0.893 | <u>0.909</u> | 0.899 | 0.901 | 0.904 | **0.916** |
|  | MSAM | 0.651 | 0.521 | 0.204 | 0.214 | 0.278 | 0.256 | 0.211 | 0.214 | 0.201 | <u>0.186</u> | 0.201 | **0.194** |
| Case6 | MPSNR | 14.10 | 22.41 | 32.84 | 30.11 | 29.11 | 29.36 | 30.53 | 32.91 | <u>32.94</u> | 31.47 | 31.45 | **34.12** |
|  | MSSIM | 0.229 | 0.624 | 0.911 | 0.895 | 0.871 | 0.878 | 0.907 | <u>0.917</u> | 0.907 | 0.914 | 0.916 | **0.925** |
|  | MSAM | 0.678 | 0.354 | <u>0.116</u> | 0.157 | 0.224 | 0.210 | 0.132 | 0.134 | 0.140 | 0.122 | 0.124 | **0.115** |

Tables II, III, and IV present the mean peak signal-to-noise ratio (MPSNR), mean structural similarity index measure (MSSIM), and mean spectral angle mapper (MSAM) values obtained for all the denoising methods on the Pavia City Center, Indian Pines, and Washington DC Mall datasets. The best results for each case are highlighted in bold and the second-best results are underlined. It can be seen that the proposed LRTDAHL method exhibits a better overall performance than the other comparison methods in removing mixed noise, particularly in the presence of strong stripe noise or specific stripes. In the absence of stripe noise, the proposed method is very close to the optimal results. It is worth noting that BM4D performs the worst in all cases, due to its disregard for the spectral correlation of the HSI and sparse noise. In most cases, the matrix factorization based LRTV outperforms the tensor-based methods of LRTDTV and 3DlogTNN in terms of performance, as the matrix formulation process can destroy the spatial information of the band pixels. The noise modeling based NMoG method and the non-convex low-rank approximation based FGSLR method exhibit slight advantages or comparability to the proposed method in cases of only Gaussian noise and sparse noise, but their performance is inferior to that of LRTDAHL in the presence of stripe noise. Overall, these quantitative comparisons demonstrate the effectiveness of the proposed method in HSI denoising and stripe removal.

Fig. 3 presents a comparison of the peak signal-to-noise ratio (PSNR) and structural similarity index measure (SSIM) values for various bands in Case1, Case3, and Case5 of the simulated experiments. The results indicate that the performance of BM4D is generally poor, while the performance of LRTV and LRTDTV is not very robust. Although the proposed method does not show significant advantages over the state-of-the-art methods in Case1, it exhibits considerable advantages in almost all the bands when high-intensity stripe noise and specific stripes are present.

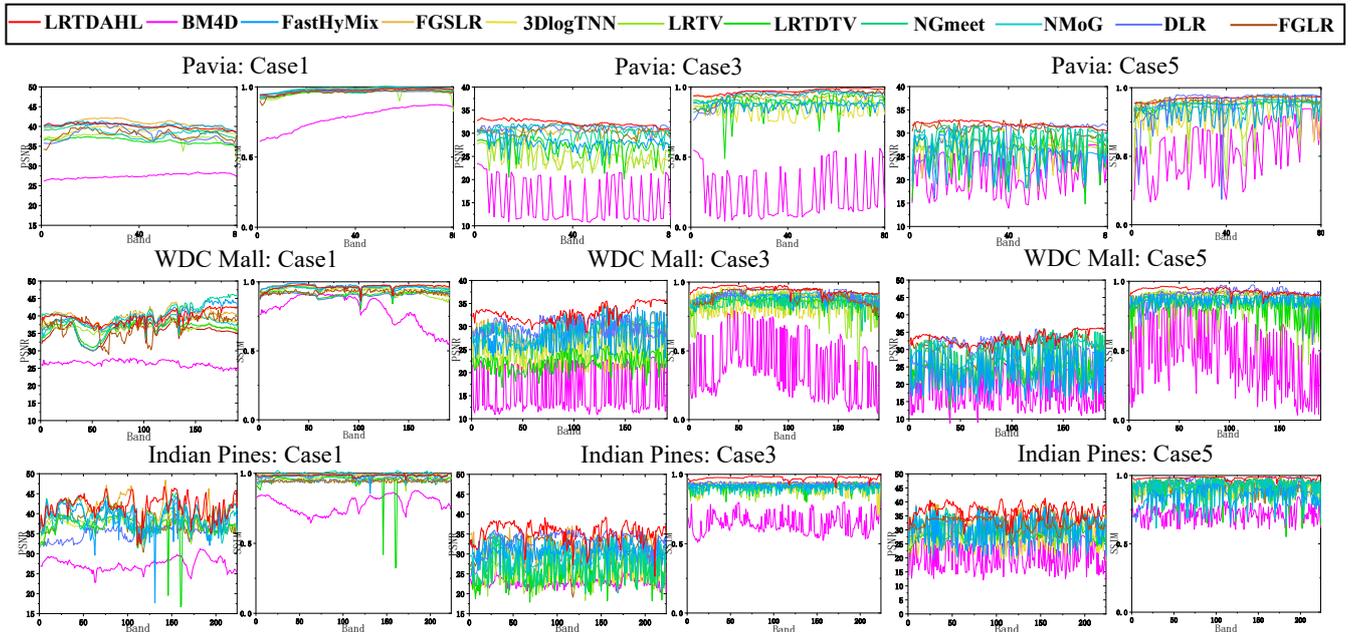

Fig. 3. PSNR and SSIM values of each band of the restored results obtained by the different methods.

3) *Qualitative Comparison:* Further performance comparisons of the nine comparison methods were made, and their results were visually analyzed. Specifically, the Pavia City Center, Indian Pines, and Washington DC Mall datasets were used to evaluate the denoising performance of each method under the Case1, Case3, Case4, Case5, and Case6 scenarios, as shown in Figs. 5, 6, and 7. The subregions marked in each subplot are enlarged and framed in red boxes for better visual clarity. It can be observed that the proposed method achieves comparable visual results to the other state-of-the-art methods in the absence of stripe noise. However, the proposed method demonstrates the best visual performance under the high stripe noise intensity and special stripe noise cases. On the other hand, the matrix-based methods, such as LRTV, and the tensor-based methods, such as LRTDTV and 3DlogTNN, leave considerable residual stripe noise. The noise modeling based methods, such as NMoG, and the non-convex low-rank approximation based methods, such as FGSLR, achieve reasonable denoising performances in some cases, but noise residuals and loss of details can still be observed in the enlarged boxes. The methods of DLR and FGLR, which are based on modeling low-rank matrices with stripe noise, can effectively remove a significant proportion of the stripe noise. However, in certain situations, they can struggle with handling subtle stripe artifacts and recovering fine image details. In summary, the proposed LRTDAHL method outperforms the other methods in removing mixed noise containing stripe noise, while also preserving the structure and local details of the clean HSIs.

To assess the spectral recovery capability of each method, Fig. 4 illustrates the clean spectra (blue curves) and denoised spectra (red curves) of pixels in the simulated datasets. It is evident that the mixed noise has disrupted the spectral characteristics of the pixels, and BM4D is incapable of restoring the original spectral curves. LRTV, LRTDTV, 3DlogTNN, and FastHyMix can recover most of the frequency bands, but sharp noise persists in some bands. NMoG, NGmeet, and FGSLR yield results with significant noise in some bands. Overall, the proposed LRTDAHL method can effectively restore the spectral characteristics of the pixels.

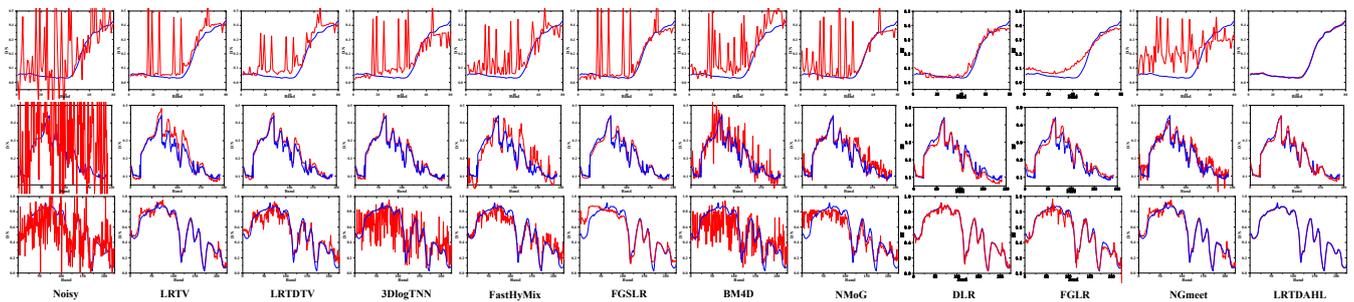

Fig. 4. Spectral curves of the restored results obtained by the different methods. The red curves are the denoising results and the blue curves are the original spectral curves. The first row shows the results obtained at spatial location (134,76) of the Pavia City Center dataset under Case6. The second row shows the results obtained at spatial location (107,214) of the Washington DC Mall dataset under Case6. The last row shows the results obtained at spatial location (30,30) of the Indian Pines dataset under Case6.

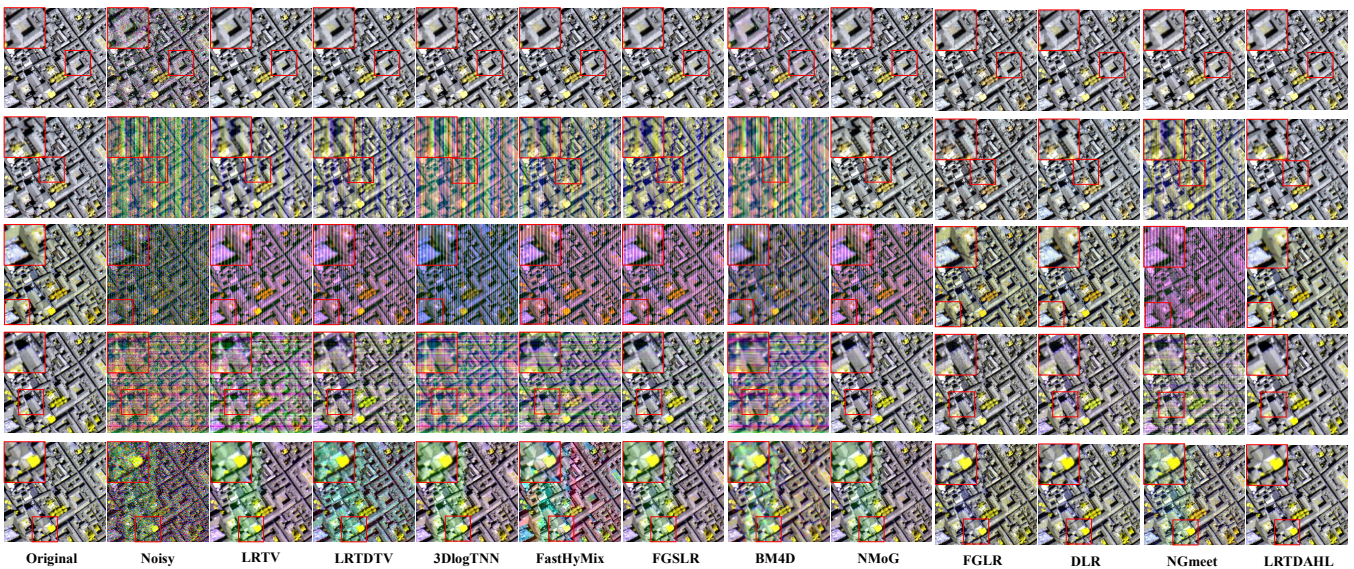

Fig. 5. Denoising false-color results obtained on the Pavia City Center dataset by the different methods. From the first row to the last row are Case1, Case3, Case4, Case5, and Case6, respectively.

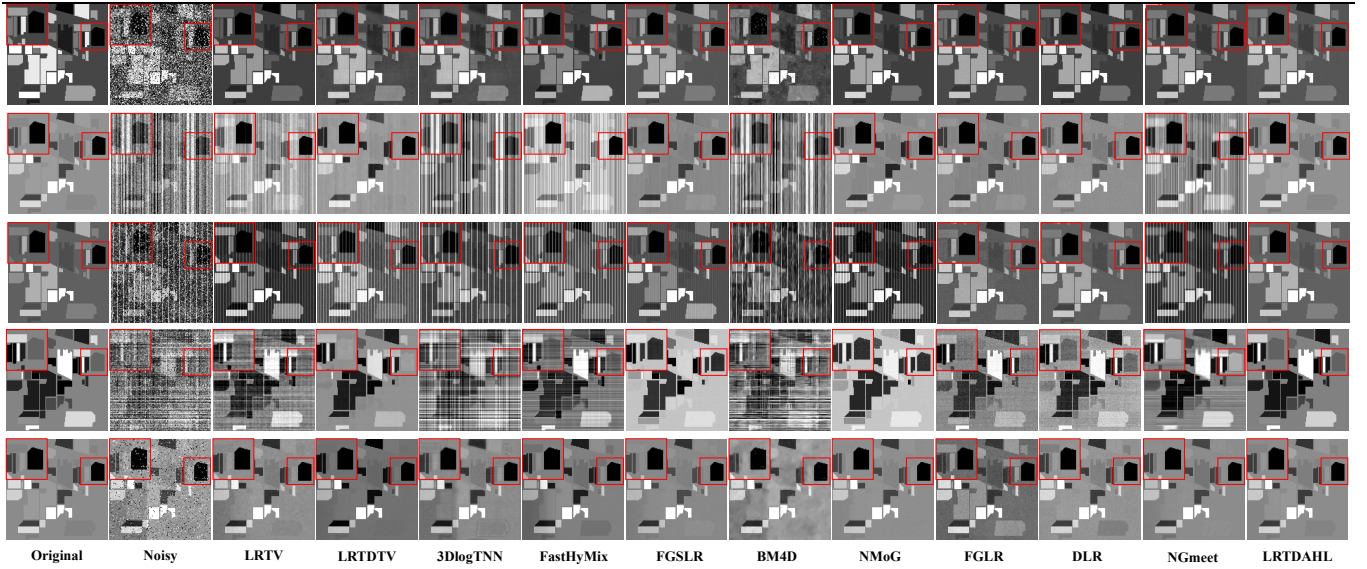

Fig. 6. Denoising results obtained on the Indian Pines dataset by the different methods. From the first row to the last row are Case1, Case3, Case4, Case5, and Case6, respectively.

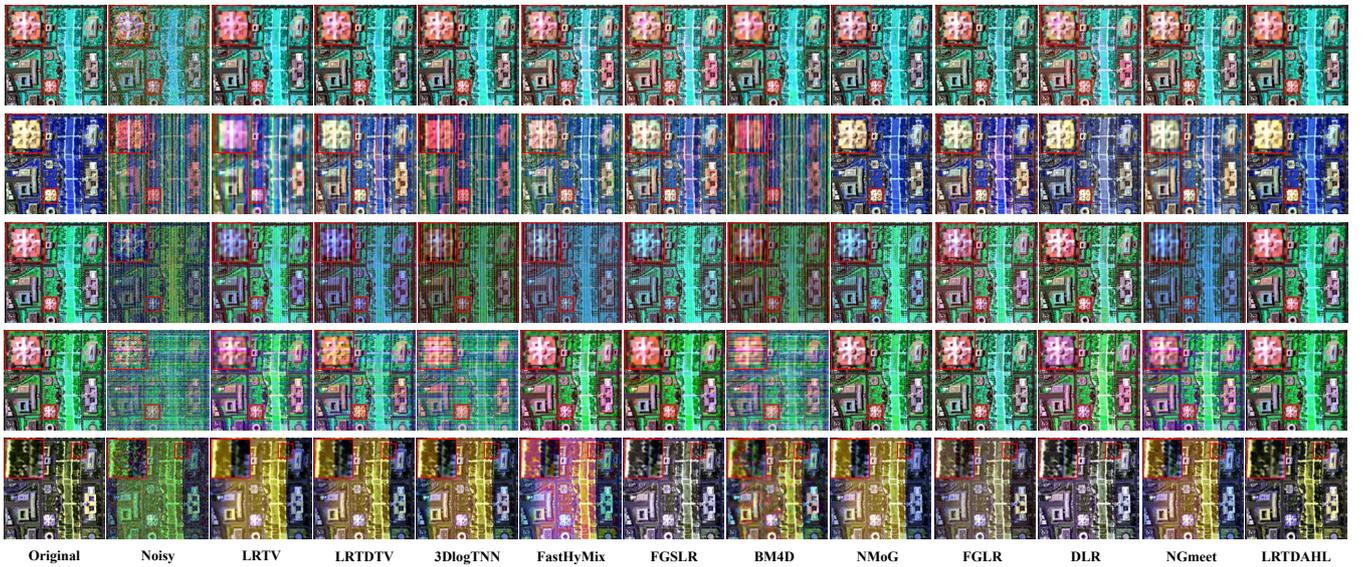

Fig. 7. Denoising false-color results obtained on the Washington DC Mall dataset by the different methods. From the first row to the last row are Case1, Case3, Case4, Case5, and Case6, respectively.

## B. Real-Data Experiments

The effectiveness of the proposed model was tested using three real datasets, namely, the Hyperspectral Digital Imagery Collection Experiment (HYDICE) Urban dataset, an Earth Observing-1 (EO-1) Hyperion dataset, and a Gaofen-5 (GF-5) dataset. Prior to applying the denoising model, the HSIs were normalized band-by-band to the range of [0,1].

*1) HYDICE Urban Dataset*: The HYDICE Urban dataset is a publicly available and widely used dataset in hyperspectral remote sensing data recovery experiments. It was acquired by the HYDICE sensor and contains a total of 210 bands ranging from 0.4 to 2.4μm in the visible and near-infrared spectral range. The dataset is severely contaminated by Gaussian noise, dead lines, stripe noise, water absorption noise, atmospheric effects, and some unknown noise.

Figs. 8 and 9 present the denoised images of the 108th and 150th bands of the HYDICE Urban dataset obtained using the different methods. The images are shown before and after denoising, with local zoomed-in details displayed in the red boxes below each image. The original images are severely corrupted by mixed noise, including dense stripes, Gaussian noise, and impulse noise. It can be observed that the LRTV, 3DlogTNN, and BM4D methods only remove a small amount of the mixed noise, and cannot eliminate the dense noise near the image edges and the wide stripe noise, which can cause significant image distortion. The LRTDTV and FastHyMix methods exhibit an excellent performance in removing the Gaussian and sparse noise, but perform poorly in removing the stripe noise, and the zoomed-in details reveal the presence of significant stripe noise even after denoising. Visually, FGSLR, NMoG, DLR, and the proposed LRTDAHL method demonstrate the best performances, effectively removing the mixed noise while preserving the spa-tial details. The narrow distribution of the noise-contaminated bands, the concentrated contamination intensity, and the lack of prominent striping noise contamination contribute to the relatively easier separation between noise and clean signals in this dataset, which could be a potential explanation for the comparable results obtained by the proposed method and the state-of-the-art approaches.

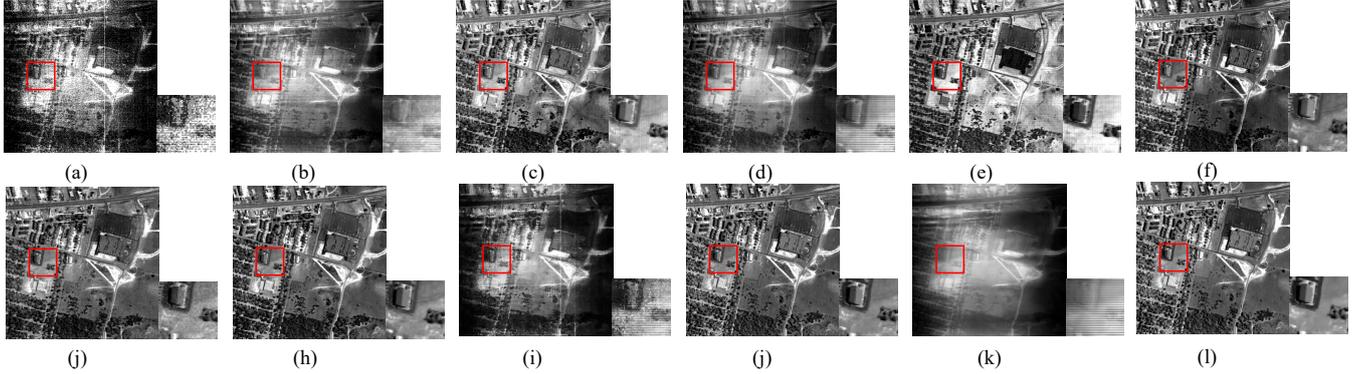

Fig. 8. Denoised results for band 108 of the HYDICE Urban dataset obtained by the different methods. (a) Original. (b) LRTV. (c) LRTDTV. (d) 3DlogTNN. (e) FastHyMix. (f) DLR. (g) FGLR. (h) FGSLR. (i) BM4D. (j) NMoG. (k) NGmeet. (l) LRTDAHL.

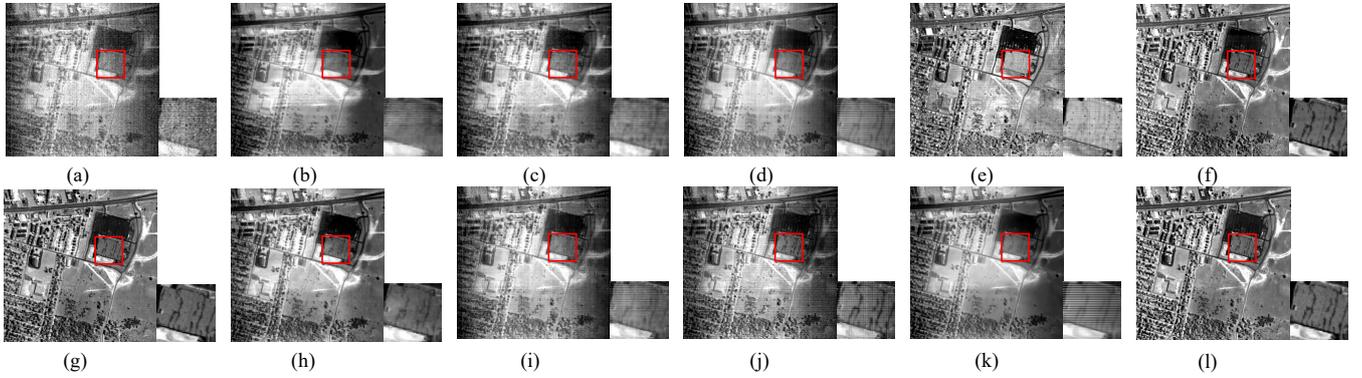

Fig. 9. Denoised results for band 139 of the HYDICE Urban dataset obtained by the different methods. (a) Original. (b) LRTV. (c) LRTDTV. (d) 3DlogTNN. (e) FastHyMix. (f) DLR. (g) FGLR. (h) FGSLR. (i) BM4D. (j) NMoG. (k) NGmeet. (l) LRTDAHL.

2) *EO-1 Hyperion Dataset*: This dataset was obtained by the Hyperion hyperspectral imaging spectrometer, spanning 400–2500nm across a total of 242 bands, with a spectral resolution of 10nm. However, the dataset is severely corrupted by various forms of noise, including Gaussian noise, dead pixels, stripe noise, water absorption noise, atmospheric noise, and other unidentified noise sources.

Figs. 10 and 11 present the restoration results for the EO-1 dataset obtained through the different methods. The original image is contaminated by stripe noise, Gaussian noise, sparse noise, and dead pixels. The BM4D method exhibits weak noise removal capabilities. Although NGmeet performs well in noise reduction, it generates excessive smoothing and loses local details. The results obtained through DLR and FGLR processing still exhibit subtle remnants of stripe artifacts. The other comparison methods can eliminate most of the Gaussian and sparse noise, but stripe noise and noticeable dead lines remain. Overall, the proposed LRTDAHL method demonstrates the best performance in terms of noise removal.

3) *GF-5 Dataset:* The GF-5 dataset was acquired using the Advanced Hyperspectral Imager (AHSI) instrument, and consists of a total of 330 bands ranging from 400 to 2500 nm. This GF-5 short-wave infrared (SWIR) dataset is contaminated by mixed noise, including stripe noise, Gaussian noise, sparse noise, and dead pixels.

From Fig. 12 and Fig. 13, it can be observed that the BM4D method exhibits a relatively weak performance. The LRTV, LRTDTV, FastHyMix, and FGSLR methods can remove most of the noise, but still leave significant stripe artifacts and produce over-smoothing. The NMoG, FGLR, and NGmeet methods demonstrate a better performance, but still leave stripe noise artifacts and cause spectral distortion. Overall, the proposed LRTDAHL method and the DLR method can effectively remove the mixed noise and demonstrate the best performance.

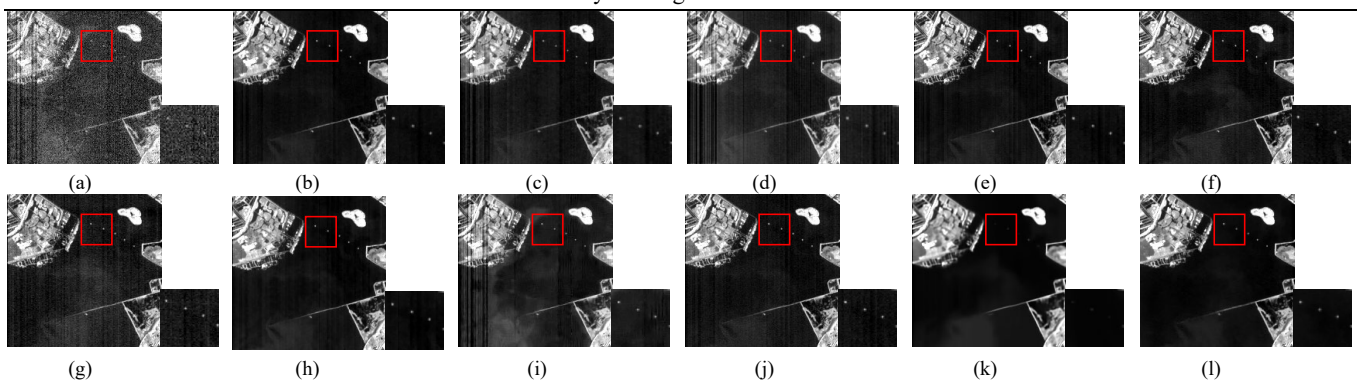

Fig. 10. Denoising results for band 130 of the EO-1 dataset obtained by the different methods. (a) Original. (b) LRTV. (c) LRTDTV. (d) 3DlogTNN. (e) FastHyMix. (f) DLR. (g) FGLR. (h) FGSLR. (i) BM4D. (j) NMoG. (k) NGmeet. (l) LRTDAHL.

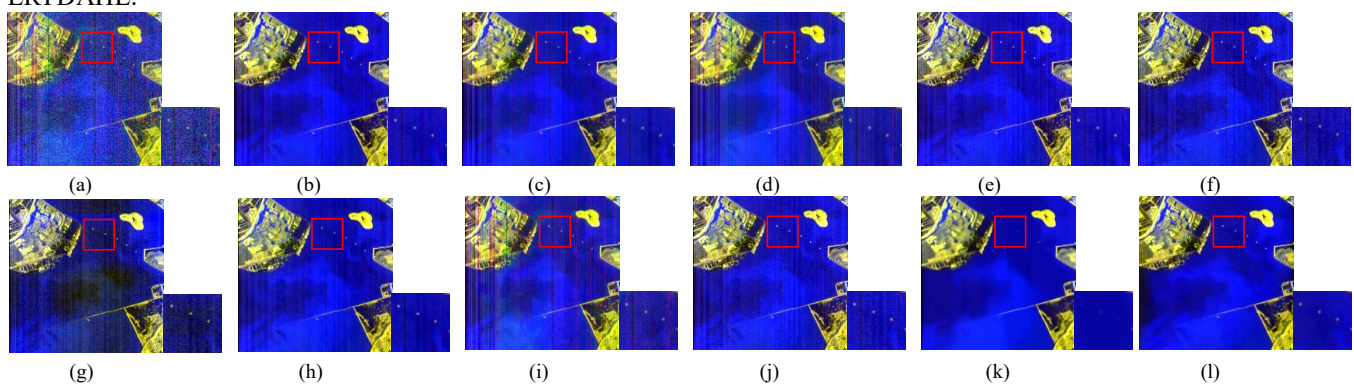

Fig. 11. Denoised false-color results for the EO-1 dataset obtained by the different methods. (a) Original. (b) LRTV. (c) LRTDTV. (d) 3DlogTNN. (e) FastHyMix. (f) DLR. (g) FGLR. (h) FGSLR. (i) BM4D. (j) NMoG. (k) NGmeet. (l) LRTDAHL.

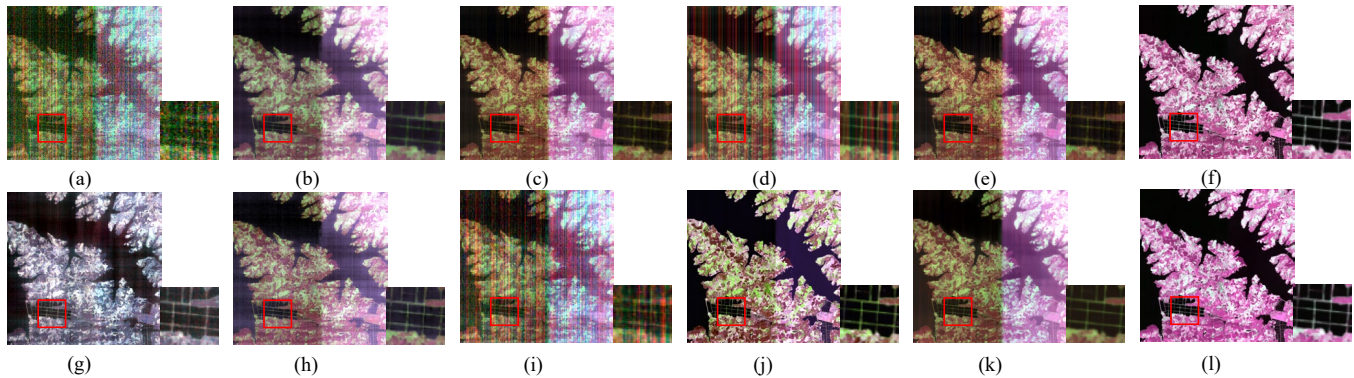

Fig. 12. Denoised false-color results for the GF5 dataset obtained by the different methods. (a) Original. (b) LRTV. (c) LRTDTV. (d) 3DlogTNN. (e) FastHyMix. (f) DLR. (g) FGLR. (h) FGSLR. (i) BM4D. (j) NMoG. (k) NGmeet. (l) LRTDAHL.

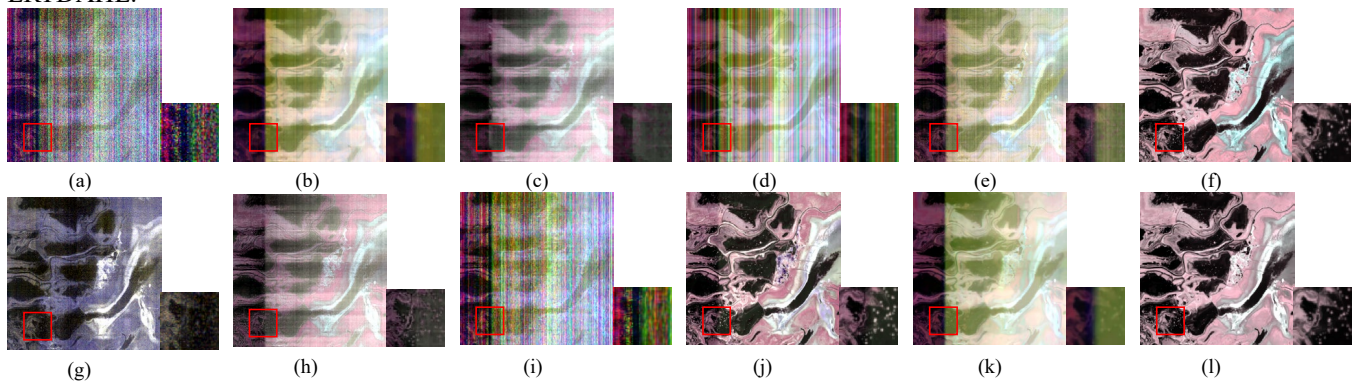

Fig. 13. Denoised false-color results for the GF5 dataset obtained by the different methods. (a) Original. (b) LRTV. (c) LRTDTV. (d) 3DlogTNN. (e) FastHyMix. (f) DLR. (g) FGLR. (h) FGSLR. (i) BM4D. (j) NMoG. (k) NGmeet. (l) LRTDAHL.

*C. Discussion*

*1) Parameter Analysis:* The proposed LRTDAHL method's performance in denoising HSIs is affected by the selection of regularization parameters $\lambda_1$ and $\lambda_2$ and rank values $r_1$ and $r_2$. In this study, we investigated the influence of these parameters on the proposed method's performance.

The impact of parameter variations in the denoising of the simulated noisy images from Case1 to Case6 was investigated. Specifically, the effect of $\lambda_1$ and $\lambda_2$ on the denoising results was analyzed. $\lambda_1$ is the contribution parameter for the spatial-spectral TV constraint, while $\lambda_2$ is the regularization parameter for sparse noise. The proposed method was tested with varying $\lambda_1$ and $\lambda_2$ values, which were selected from the ranges of [0, 0.001, 0.002, 0.003, 0.004, 0.005, 0.006, 0.007] and [0, 0.01, 0.02, 0.03, 0.04, 0.05, 0.06, 0.07], respectively.

The presented model's performance on the simulated Pavia City Center dataset with varying values of parameters $\lambda_1$ and $\lambda_2$ is illustrated in Fig. 15. In Case1, from Fig. 14(a), it can be seen that when the noise types are limited to Gaussian noise and sparse noise, the proposed method achieves optimal MPSNR values in the vicinity of $\lambda_1$ ranging from 0.001 to 0.003 and $\lambda_2$ ranging from 0 to 0.02. In Case2 to Case 6, as illustrated in Fig. 15(b)–(f), when adding stripe noise, a not very low value of $\lambda_1$ can bring better results. The proposed approach achieves optimal MPSNR values in the vicinity of $\lambda_1$ ranging from 0.002 to 0.007 and $\lambda_2$ ranging from 0.02 to 0.07. Both excessively small and large values of $\lambda_1$ lead to diminished denoising results, with $\lambda_2$ exhibiting a relatively lower sensitivity, compared to $\lambda_1$. Based on this observation, $\lambda_1$ and $\lambda_2$ were set to 0.002 and 0.02, respectively, in both the simulated and real-data experiments conducted in this study.

*2) Effectiveness of the Stripe Tensor Decomposition Term:* In this section, we discuss the superiority of the stripe tensor decomposition term in the LRTDAHL model.

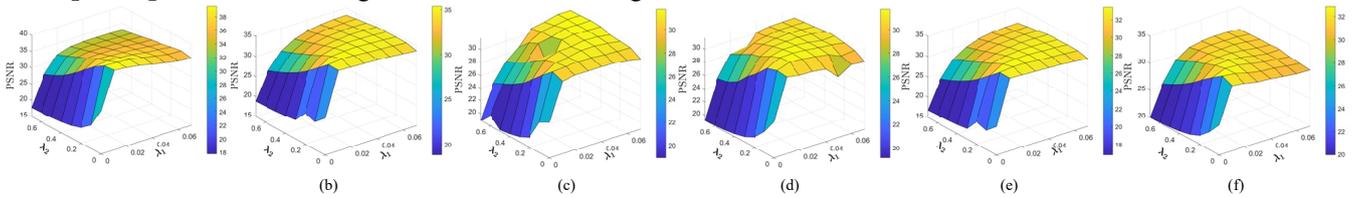

Fig. 14. Sensitivity analysis for parameters $\lambda_1$ and $\lambda_2$ in the simulated experiments. (a) Case1. (b) Case2. (c) Case3. (d) Case4. (e) Case5. (f) Case6.

The quantitative results of the experiments on the simulated Pavia City Center dataset are shown in Fig. 15(a)–(b), where the original images were contaminated with noise similar to Case 2 to Case 6. We compared the performance of LRTDAHL with stripe tensor decomposition and stripe matrix decomposition separately, and it can be observed that the performance is worst when the stripe noise is not modeled separately. Using tensor decomposition to characterize stripes performs better than using matrix decomposition, as tensor decomposition can simultaneously encode the high correlation of the stripe components in the two spatial directions and the correlation in part of the spectral direction, as opposed to the per-band matricization. Fig. 15(c)–(e) show the image results with and without stripe tensor decomposition, and it can be observed that, without applying the low-rank tensor term for stripes, it is impossible to completely separate the periodic stripes in the spatial direction from the image components. Fig. 15(f)–(i) show the decomposition components of the GF-5 dataset in band 180 after denoising with LRTDAHL, and it can be visually observed that the proposed method can separate the clean imagery, sparse noise, Gaussian noise, and stripe noise, which potentially demonstrates the effectiveness of the two tensor decomposition terms in LRTDAHL.

*3) Effectiveness of the Adaptive Hyper-Laplacian Term:* In this section, the effectiveness of the adaptive hyper-Laplacian term in the LRTDAHL model is discussed.

As is well known, different HSIs often exhibit varying gradient distributions, which results in the density functions of the optimal hyper-Laplacian distributions they conform to being dissimilar. In this context, Fig. 16 presents the experimental results obtained on the simulated Pavia City Center dataset, where the original images were contaminated with the same noise as in Case2. Specifically, Fig. 16(a)–(c) depict the quantitative results of the denoising performance after varying $p_h$, $p_w$, and $p_p$ while keeping the other parameters constant. The proposed LRTDAHL method obtains a good result with its estimated $p$ values, with the results consistently near the optimal values. In addition, Fig. 16(e)–(g) present the denoising results for the false-color images obtained with different $p_h$ values while keeping the other parameters constant. The value $p_h$ = 0.642 is the result of the adaptive estimation. It can be observed that excessively large $p$ values lead to residual noise, while excessively small $p$ values lead to overly smooth results with lost image details. In contrast, the proposed method's estimated $p$ values exhibit a good performance.

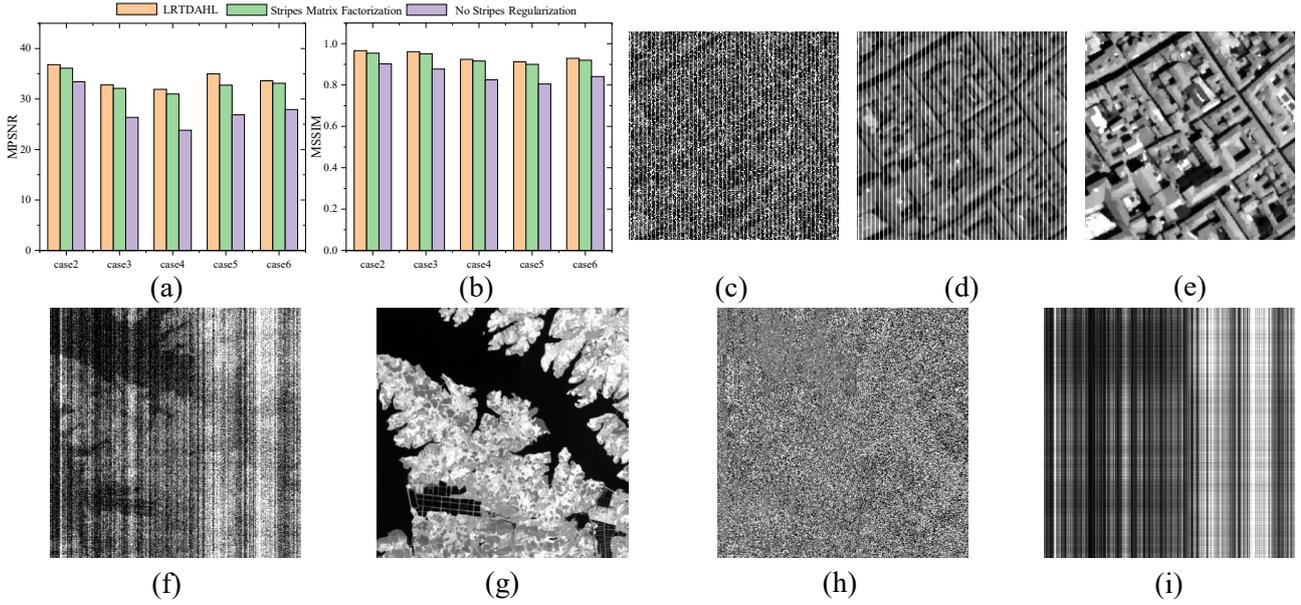

Fig. 15. Effectiveness of the stripe tensor decomposition. The results for Case2 to Case6 for LRTDAHL with stripe matrix decomposition and no stripe tensor decomposition. (a) MPSNR. (b) MSSIM. The results for band 40 in Case4. (c) Noisy. (d) No stripe tensor decomposition. (e) LRTDAHL result. The results for band 180 in the GF-5 dataset. (f) Original. (g) LRTDAHL result. (h) Gaussian and sparse noise. (i) Stripe noise.

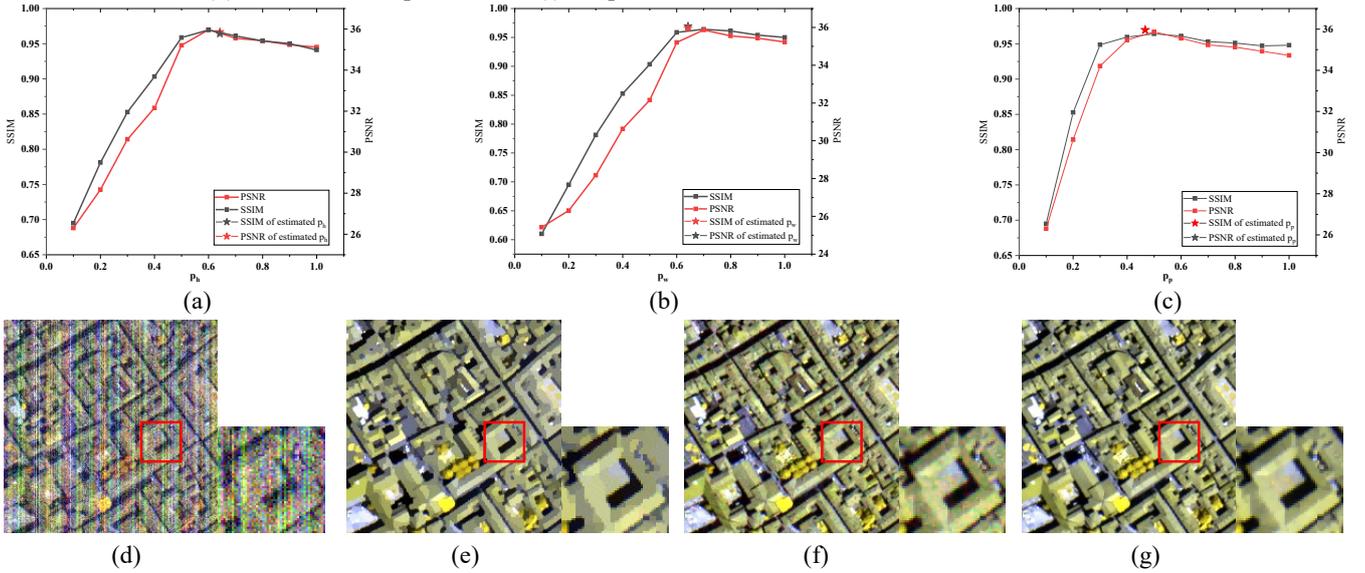

Fig. 16. Effectiveness analysis for the adaptive hyper-Laplacian prior. (a) Performance with different $p_h$ values for the hyper-Laplacian prior (estimated $p_h$ = 0.642). (b) Performance with different $p_w$ values for the hyper-Laplacian prior (estimated $p_w$ = 0.684). (c) Performance with different $p_p$ values for the hyper-Laplacian prior (estimated $p_p$ = 0.485). (d) Noisy false-color image of Case5. (e) False-color image for $p_h$ = 0.3. (f) False-color image for $p_h$ = 1.0. (g) False-color image for $p_h$ = 0.642.

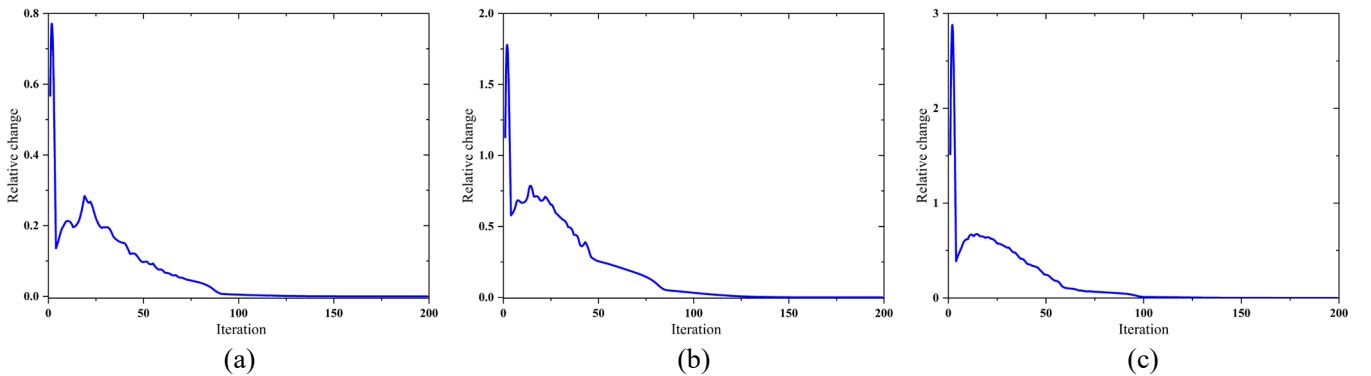

Fig. 17. Relative change values of LRTDAHL with respect to the iteration number. (a) HYDICE Urban dataset. (b) EO-1 dataset. (c) GF-5 dataset.

3) *Convergence of the LRTDAHL Solver*: It is evident that the proposed model (11) is a non-convex optimization problem, and non-convex optimization problems are difficult to find global optimal solutions for and prove their convergence. Therefore, we introduce the ALM method to optimize the proposed model. Fig. 17 shows the evolution curve of the relative change with iteration number in the real-data experiments. It can be observed that the relative change converges to zero when the iteration number reaches a high value, indicating that the convergence of LRTDAHL can be guaranteed.

## IV. Conclusion

In this paper, we have proposed the tensor-based LRTDAHL method for HSI denoising and destriping. Specifically, we utilize tensor low-rank decomposition to characterize the spatial-spectral correlation of the HSI and stripe noise and introduce adaptive hyper-Laplacian regularization to further encode the spatial-spectral information of the HSI gradients. We then introduce the ADMM algorithm to efficiently optimize the proposed model.

Extensive simulation and real-data experiments were conducted, and the results clearly demonstrated that the proposed LRTDAHL method outperformed the other classical and state-of-the-art methods in both the visual and quantitative evaluations. This is because the introduced stripe tensor decomposition term helps separate stripes from the clean HSI, and the adaptive hyper-Laplacian regularization term is better able to preserve the details of the image, compared to the classical TV and Laplacian regularization priors, while removing the residual artifacts.